\documentstyle[12pt,fullpage,epsfig]{article}

\def\be{\begin{equation}}
\def\ee{\end{equation}}
\def\bea{\begin{eqnarray}}
\def\eea{\end{eqnarray}}

\def\gsim{\:\raisebox{-0.5ex}{$\stackrel{\textstyle>}{\sim}$}\:} 
\begin{document}
\begin{flushright}
TIFR/TH/00-66 \\
\end{flushright}
\bigskip
\begin{center}
{\Large{ \bf Electron and Neutron Electric Dipole Moments in the Focus Point
Scenario of SUGRA Model}} \\[1.5cm]
Utpal Chattopadhyay$^{(a)}$, Tarek Ibrahim$^{(b)}$ and D.P. Roy$^{(a)}$\\[0.75cm]
$^{(a)}$Department of Theoretical Physics \\ Tata Institute of Fundamental
Research, Mumbai - 400 005, India \\[0.5cm] 
$^{(b)}$Department of Physics, Faculty of Science\\ University of Alexandria,
Alexandria, Egypt
\end{center}
\bigskip\bigskip

\begin{abstract}

We estimate the electron and neutron electric dipole moments in the
focus point scenario of the minimal SUGRA model corresponding to
large sfermion masses and moderate to large $\tan\beta$.  There is a
viable region of moderate fine-tuning in the parameter space, around 
$\tan\beta \simeq 5$, where the experimental limits on these electric
dipole moments can be satisfied without assuming unnaturally small
phase angles.  But the fine-tuning constraints become more severe for
$\tan\beta > 10$.

\vspace{4mm}

\end{abstract}

\newpage

It has been long recognised that the experimental limits on the
electron and neutron electric dipole moments (EDM) imply stringent
constraints on the minimal supersymmetric standard model (MSSM) and in
particular the minimal supergravity (SUGRA) model \cite{b1}.  In 
order to satisfy these
limits one has to assume either unnaturally small phase angles $(<
10^{-2})$ in the model or multi-TeV superparticle masses \cite{b2}.  
More recently it has been shown by the authors of ref.~\cite{b3} 
that the problem is alleviated to a large extent by internal 
cancellation between different supersymmetric (SUSY) 
contributions to these EDMs.  
Consequently, one can
satisfy the experimental constraints on the EDMs with moderate phase
angles and moderate superparticle masses in the unconstrained version
of the MSSM \cite{b4,b5}.  However one still requires large superparticle
masses in the minimal SUGRA model \cite{b3,b6,b7}, which is undesirable for
three reasons.  It implies, i) a large fine-tuning parameter for
radiative breaking of electroweak symmetry, ii) a less viable SUSY
signal at the forthcoming colliders, and iii) a very large dark matter 
density of the universe \cite{b7}.

Recently, Feng, Matchev and Moroi \cite{b8} have pointed out that the
radiative electroweak symmetry breaking condition and hence the
resulting fine-tuning is practically independent of the universal 
soft scalar mass parameter $m_0$ in the minimal SUGRA model for moderate 
to large values of $\tan\beta (\gsim 5)$.  This is also the 
range favoured by the LEP data\cite{b9}.  This has been referred to as 
the focus point phenomenon.  It
implies that one can have a $m_0$ and hence sfermion masses of the
first two generations in the multi-TeV region without affecting the
fine-tuning parameter of electroweak symmetry breaking.  Besides, one
expects in this case an inverted hierarchy of squark masses, resulting
in a distinctive SUSY signal at the LHC from gluino production
\cite{b10,b11}.  Moreover, it has been shown to predict a dark matter density
of the universe, which is in fact in the desired range \cite{b12}.

In this paper we have calculated the electron and neutron EDMs in the
focus point scenario to see if they can be reconciled with the
corresponding experimental limits without assuming unnaturally small
phase angles.  The large mass of the first generation sfermions in
this model helps to suppress the electron and neutron EDMs.  Moreover, 
a large value of the trilinear coupling parameter $A_0$ helps to
suppress them further via a more effective cancellation between the 
different SUSY contributions.  But this is partly offset by the
increase of these EDMs with $\tan\beta$.  Thus for $\tan\beta > 
10$, one cannot satisfy the experimental limits without assuming
unnaturally small phase angles.  However, there is a viable region of
the parameter space at around $\tan\beta \simeq 5$, where the
experimental limits can be satisfied with moderate values of the phase
angles.  

In the following section we briefly discuss the focus point scenario
of the minimal SUGRA model and estimate the fine-tuning parameter over
the region of interest to the EDM calculation.  In the next section we
discuss the EDM calculation and identify the region of parameter
space, where the EDMs can be reconciled with the experimental limits
for moderate phase angles.  We shall conclude with a brief summary of
our results.
\bigskip

\noindent {\bf Focus Point and Fine-tuning:}
\medskip

\nobreak
The basic parameters of the minimal SUGRA model are
$m_0,M_{1/2},A_0,B$ and $\mu$ -- i.e. soft supersymmetry breaking
scalar and gaugino masses, trilinear and bilinear couplings, along
with the supersymmetric Higgs mass parameter \cite{b13}.  The last two can
be determined in terms of the two Higgs vacuum expectation values,
$v_1$ and $v_2$, using the two minimisation conditions.  The first
condition determines $B$ in terms of the ratio $v_2/v_1 \equiv
\tan\beta$ and the sum
\be
v^2 = v^2_1 + v^2_2 = 2m^2_Z/(g^2 + g^{\prime 2}) \simeq 175 \ {\rm
GeV}. 
\label{one}
\ee
The second condition gives
\be
{1\over2} m^2_Z = {m^2_{H_1} - m^2_{H_2} \tan^2\beta \over \tan^2
\beta - 1} - |\mu|^2 + \Delta_R,
\label{two}
\ee
where the last term comes from the radiative correction to the Higgs
potential.  This equation determines the modulus of $\mu$.

Thus for any $\tan\beta$, the naturalness of the electroweak symmetry
breaking scale requires $m^2_{H_2}$ and $|\mu|^2$ to be of the order
of $m^2_Z$, so that there is no large cancellation between these 
quantities \cite{b14}.  Since $m^2_{H_2}$ is linearly related to 
$m^2_0,M^2_{1/2}$ and ${|A_0|}^2$ via its 
renormalisation group equation (RGE), one usually 
assumes the naturalness criterion to imply each of these parameters to 
be $<1~{\rm TeV}$.  Indeed, most of the phenomenological studies 
within the minimal SUGRA model are based on this assumption.  
However, as pointed out by Feng et al \cite{b8}, for physical values of 
the top quark mass and the gauge couplings, $m^2_{H_2}$ at the 
electroweak scale becomes practically independent of its GUT scale value 
$m^2_0$ for $\tan\beta \gsim 5$. 
One can see this result from the approximate analytic solution of the
one-loop RGE for $m^2_{H_2}$ \cite{b15,b16}.  For $\tan\beta$ not too large, 
one gets while neglecting the $b$ Yukawa coupling contribution, 
\be
m^2_{H_2} \simeq m^2_0 - {3\over2} y m^2_0 + f(M_{1/2},A_0,y),
\label{three}
\ee
Here $f$ is a quadratic function of the soft parameters $M_{1/2}$ and $A_0$, 
and $y$ represents the top Yukawa coupling squared relative to its
fixed point value, i.e.
\be
y = {h^2_t \over h^2_f} = {1 + 1/\tan^2\beta \over 1 +
1/\tan^2\beta_f}.
\label{four}
\ee
The top Yukawa coupling is related to its running mass,
\be
h_t = m_t (M_t)/v \sin\beta,
\label{five}
\ee
which is related in turn to the physical top quark mass via
\be
M_t = m_t (M_t) \left[1 + \Delta_{\rm QCD} + \Delta_{\rm SUSY}\right].
\label{six}
\ee
The QCD and SUSY radiative corrections add about 6\% and 4\%
respectively to the running mass to arrive at the physical top pole
mass, $M_t = 175 \pm 5$ GeV \cite{b9}.  It is well known now that a physical
top mass of 175 GeV corresponds to the fixed point value, $\tan\beta_f
\simeq 1.5$ at the electroweak scale \cite{b16}, which defines the minimal
value of $\tan\beta$ in this model.  Such a low value of $\tan\beta$
is of course ruled out by the recent LEP limit on the lightest higgs
boson mass \cite{b9}, suggesting $\tan\beta > 2(4)$ for maximal (small) stop
mixing.  Substituting the above value of $\tan\beta_f$ in
(\ref{four}) gives
\be
y \simeq 2/3 \ {\rm for} \ \tan\beta \gsim 5.
\label{seven}
\ee
Thus over a large range of $\tan\beta$, which is also favoured by the
above mentioned LEP data, $m^2_{H_2}$ of eq.~(\ref{three}) 
at the electroweak scale is practically independent of its GUT scale value
$m^2_0$.  This is the so called focus point phenomenon, which implies
that $m_0$ can be made $> 1$ TeV without affecting the naturalness
criterion.  The corresponding squark and slepton masses of the first two
generations remain large at the electroweak scale, 
\be
m^2_{\tilde q,\tilde l} \simeq m^2_0 + O(M^2_{1/2}) > 1 \ {\rm TeV},
\label{eight}
\ee
since their RGEs are not affected by the top Yukawa coupling.
Interestingly, the focus point condition ensures that ${|\mu|}^2$ at
the electroweak scale is practically the same as its GUT scale value,
since \cite{b16}
\be
|\mu|^2 = 1.8 {|\mu_0|^2} (1 - y)^{1\over2} \simeq |\mu_0|^2.
\label{nine}
\ee

The sensitivity of the electroweak scale to the SUSY parameters are
determined from eq. (\ref{two}) in terms of the partial derivatives
\be
C_a = \left|{a \over m^2_Z} {\partial m^2_Z \over \partial a}
\right|,
\label{ten}
\ee
where $a$ denotes $m_0,M_{1/2},\mu_0$ and $A_0$.  The fine-tuning 
is defined by the largest of these quantities \cite{b14}
\be
C = {\rm max}\left\{C_{m_0},C_{M_{1/2}},C_{\mu_0},C_{A_0}\right\}.
\label{eleven}
\ee
This parameter is a plausible though not unique measure of fine-tuning. It is 
based on the sensitivity of the electroweak scale to the SUSY parameters, but
not other quantities like $m_t$\cite{b8}.

For estimating the fine-tuning parameter we have taken the radiative
electroweak symmetry breaking code of ref. \cite{b17}, which uses two-loop
RGEs along with two-loop QCD correction to the top quark mass of
eq.~(\ref{six}); and added the one-loop SUSY correction to the latter
following ref. \cite{b18}.  The radiative correction to the Higgs potential
in (\ref{two}) is evaluated using the complete one-loop result \cite{b19}.

We have computed the fine-tuning parameter $C$ in the $(m_0 - A_0)$ planes
of Figs.~(1a) to (1d) for fixed values of $M_{1/2}$. 
Figs. (1a) and (1b) show the contour plots 
of $C$ for $M_{1/2} = 300$ and 500 GeV at $\tan\beta = 5$, 
while Figs. (1c) and (1d) show the analogous plots at $\tan\beta = 10$.
The phases in these figures correspond to $\phi_\mu=0$, 
and $\phi_{A_0}=0$ and $\pi$, for the upper and 
the lower half regions of the contours respectively. 
In general, $C$ is very weakly sensitive to the 
phase $\phi_\mu$ as long as $\phi_\mu$ is in a range  
which satisfies the EDM constraints in a broad region of parameter space.
On the other hand, it is modestly sensitive to the phase $\phi_{A_0}$
as can be seen 
by comparing the $\phi_{A_0}=0$ and $\pi$ parts of the contours 
in Figs. (1a) to (1d). 

We see that for moderate values of $|A_0|$ 
( $<2000$~GeV ) the fine-tuning 
parameter increases appreciably with $M_{1/2}$, but it is 
effectively independent of $m_0$ at fixed $M_{1/2}$ and $A_0$.  
Figs. (1a) and (1b) indicate that, for contours with $\phi_{A_0}=0$, 
one can go up from $A_0 \simeq m_0 \simeq 0$ to
$m_0 \simeq 2000$ GeV and $A_0 \simeq 1500$ GeV without paying any
appreciable price in terms of fine-tuning.  Finally, these figures
also show that there is only a marginal improvement of the fine-tuning 
parameter in increasing $\tan\beta$ from 5 to 10.
\bigskip

\noindent {\bf EDMs of Electron and Neutron:}
\medskip

\nobreak
The EDM of an elementary fermion (electron or quark) is the coefficient $d^f$
of the effective Lagrangian
\be
{\cal L}_E = {-i \over 2} d^f \bar\psi \sigma_{\mu\nu} \gamma_5 \psi
F^{\mu\nu}, 
\label{twelve}
\ee
which has the nonrelativistic limit $d_f \psi_A^\dagger \vec\sigma
\cdot \vec E \psi_A$, $\psi_A$ being the large component of the Dirac
field.  Fig.~(2) shows the one-loop contributions to the effective
Lagrangian of eq.~(\ref{twelve}) in the MSSM, coming from the chargino and 
the neutralino exchanges, along with the gluino exchange in the case of
quark.  Denoting the generic interaction Lagrangian by
\be
-{\cal L}_{int} = \sum_{ik} {\bar\psi}_f (K_{ik} P_L + L_{ik} P_R) \psi_i
 \phi_k + {\rm H.C.},
\label{thirteen}
\ee
the one-loop EDM is given by \cite{b3}
\be
d^f = \sum_{ik} {m_i \over (4\pi)^2 m^2_k} Im(K_{ik} L^\star_{ik})
\left[Q_i A\left({m^2_i \over m^2_k}\right) + Q_k B\left({m^2_i \over
m^2_k}\right)\right],
\label{fourteen}
\ee
where $P_{L,R} = (1 \mp \gamma_5)/2$ and 
\be
A(r) = {1 \over 2(1-r)^2} \left(3 - r + {2\ell n r \over 1-r}\right),
\ B(r) = {1 \over 2(1-r)^2} \left(1 + r + {2r\ell n r \over 1 -
r}\right).
\label{fifteen}
\ee
Here $Q$ denotes electric charge.  The $Q_i$ and $Q_k$ terms in
(\ref{fourteen}) correspond to the diagrams with photon coupling to
the chargino $\chi^\pm_i$ and the sfermion $\tilde f_k$ respectively.

The presence of CP violating phases in the minimal SUGRA model is
responsible for a nonzero imaginary part for the product $K_{ik}
L^\star_{ik}$ in (\ref{fourteen}).  If one neglects sfermion flavour
mixing to avoid large flavour changing neutral current effects, then
there are two independent physical CP violating phases in this model
\cite{b20}.  They can be chosen to be the phases of $\mu$ and $A_0$, namely 
$\phi_\mu$ and $\phi_{A_0}$, while $M_{1/2}$ and $B \mu$ are 
chosen to be real~\cite{b3,b6,b7}.  The reality of $B \mu$ ensures that the 
Higgs vacuum expectation values and the resulting $\tan\beta$ are real.  
Following the renormalisation group equation of $\mu$ 
one may note that the phase of $\mu$ is scale independent.

The form of the effective Lagrangian of eq.~(\ref{twelve}) requires different
chiralities of the initial and the final state fermions, as indicated
in Fig.~(2) and eq.~(\ref{fourteen}).  For the gluino exchange
contribution, this comes from the chirality flip of the sfermion via
the $L$-$R$ mixing term in its mass squared matrix.  For the chargino
exchange contribution, this is accomplished via gaugino-higgsino mixing
in the $\chi^\pm$ mass matrix, while the sfermion preserves its
chirality.  The neutralino exchange receives contribution from both 
of these sources.  Since both the $L$-$R$ mixing sfermion mass squared
term and the higgsino-sfermion-fermion coupling are proportional to
$m_f$, all the contributions are proportional to the external fermion
mass.  Another consequence of the chirality flip is the explicit
proportionality of the contributions to the exchanged fermion mass $m_i$
in eq.~(\ref{fourteen}).

The gluino exchange contribution to the quark EDM is given by
\be
d^q_{\tilde g} = -{2e\alpha_s \over 3\pi} \sum^2_{k=1} Im(D^q_{2k}
D^{q^\star}_{1k}) {m_{\tilde g} \over M^2_{\tilde q_k}} Q_{\tilde q}
B\left({m^2_{\tilde g} \over M^2_{\tilde q_k}}\right),
\label{sixteen}
\ee
where $D^q$ is the $L$-$R$ mixing matrix for the squark $\tilde q$,
which diagonalises the corresponding $M^2_{\tilde q}$ \cite{b3}. 
\be
Im(D^q_{21} D^{q^\star}_{11}) = -Im(D^q_{22} D^{q^\star}_{12}) = {m_q
\over M^2_{\tilde q_1} - M^2_{\tilde q_2}} (|A_q| \sin \phi_q + |\mu|
\sin \phi_\mu R_q),
\label{seventeen}
\ee
where
\be
R_u = \cot\beta, \ R_d = \tan\beta,
\label{eighteen}
\ee
and $\phi_q$ is the phase of $A_q$ at the electroweak scale.  For the
first generation of fermions, the magnitudes and phases of the $A$ parameters 
at the electroweak scale are close to those of $A_0$ at
large $A_0$, since \cite{b6}
\bea
A_u &\simeq& (1 - 0.5 y) A_0 - 2.8 M_{1/2} \nonumber \\[2mm] 
A_d &\simeq& A_0 - 3.6 M_{1/2} \nonumber \\[2mm] 
A_e &\simeq& A_0 - 0.7 M_{1/2}.
\label{nineteen} 
\eea

The chargino exchange contributions to the EDMs are given by 
\bea
d^u_{\chi^+} &\simeq& {-e \alpha m_u \over 4\sqrt{2} \pi m_W \sin^2
\theta_W \sin\beta} \sum^2_{i=1} Im(V^\star_{i2} U^\star_{i1})
{m_{\chi^+_i} \over M^2_{\tilde d_1}} \left[A\left({m^2_{\chi^+_i}
\over M^2_{\tilde d_1}}\right) - {1\over 3} B\left({m^2_{\chi^+_i}
\over M^2_{\tilde d_1}}\right)\right], 
\label{twenty} \\[2mm] 
d^d_{\chi^+} &\simeq& {-e\alpha m_d \over 4\sqrt{2} \pi m_W \sin^2
\theta_W \cos\beta} \sum^2_{i=1} Im(U^\star_{i2} V^\star_{i1})
{m_{\chi^+_i} \over M^2_{\tilde u_1}} \left[-A\left({m^2_{\chi^+_i}
\over M^2_{\tilde u_1}}\right) + {2 \over 3} B\left({m^2_{\chi^+_i}
\over M^2_{\tilde u_1}}\right)\right],
\label{twentyone} \\[2mm] 
d^e_{\chi^+} &\simeq& {e\alpha m_e \over 4\sqrt{2} m_W \sin^2
\theta_W \cos\beta} \sum^2_{i=1} Im(U^\star_{i2} V^\star_{i1})
{m_{\chi^+_i} \over M^2_{\tilde \nu_e}} A\left({m^2_{\chi^+_i} \over
M^2_{\tilde\nu_e}} \right),
\label{twentytwo}
\eea
where $\tilde u_1,\tilde d_1$ refer to the dominantly left-handed
squark mass eigenstates.  In our numerical analysis we have also included
the small contributions from 
the terms with ${\tilde u}_2$, ${\tilde d}_2$.
Here, $U$ and $V$ are the gaugino-higgsino
mixing matrices, which diagonalise the chargino mass matrix.  Explicit
expression for the $U$ and $V$ matrices are given in ref.~\cite{b3} in terms of
$M_{1/2}$, $\tan\beta$, $|\mu|$ and $\phi_\mu$.  We shall simply note
here that each of the coefficients $Im(U^\star_{i2} V^\star_{i1})$
and $Im(V^\star_{i2} U^\star_{i1})$, is proportional to $|\mu|
\sin\phi_\mu$.  Consequently,
\be
d^{u,d,e}_{\chi^+} \propto |\mu| \sin\phi_\mu.
\label{twentythree}
\ee

The neutralino exchange contributions to the EDMs can be collectively
expressed as, 
\be
d^f_{\chi_0} = {e\alpha \over 4\pi\sin^2\theta_W} \sum^2_{k=1}
\sum^4_{i=1} Im(\eta^f_{ik}) {m_{\chi^0_i} \over M^2_{\tilde f_k}} Q_{\tilde f}
B\left({m^2_{\chi^0_i} \over M^2_{\tilde f_k}} \right),
\label{twentyfour}
\ee
where
\bea
\eta^f_{ik} &=& \left[-\sqrt{2} \left\{\tan\theta_W (Q_f - T_{3f})
N_{1i} + T_{3f} N_{2i}\right\} D^{f^\star}_{1k} - \kappa_f N_{bi}
D^{f^\star}_{2k}\right] \nonumber \\[2mm] 
&& \times \left(\sqrt{2} \tan\theta_W Q_f N_{1i} D^f_{2k} - \kappa_f N_{bi}
D^f_{1k} \right),
\label{twentyfive}
\eea
with
\be
\kappa_u = {m_u \over \sqrt{2} m_W \sin\beta}, \ \kappa_{d,e} = {m_{d,e} \over
\sqrt{2} m_W \cos\beta}
\label{twentysix}
\ee
and $b = 4(3)$ for $u(d,e)$.  The $D^f$ are the $L$-$R$ mixing
matrices for the sfermion $\tilde f$, which occurred earlier in 
eq.~(\ref{sixteen}).  Explicit expression for its matrix elements are
given in ref.~\cite{b3} in terms of $|A_f|$, $|\mu|$, $\phi_f$ and $\phi_\mu$.
Finally, $N$ is the $4 \times 4$ unitary matrix, 
diagonalising the neutralino mass matrix, which is evaluated numerically.

The main contributions to the EDM of quarks come from chargino and
gluino exchanges, while neutralino exchange contribution is relatively
small.  They are related to the neutron EDM via the nonrelativistic
quark model relation \cite{b21},
\be
d^n = {1\over3} \left[4d^d - d^u\right] \eta_E,
\label{twentyseven}
\ee
where $\eta_E = 1.53$ is a QCD correction factor for evolving down the
quark EDMs from the electroweak to the hadronic scale \cite{b3,b22}.

There are two other contributions to the neutron EDM, arising from the
quark chromoelectric dipole moment and the gluonic dimension-six
operator, which are defined by the effective Lagrangians
\be
{\cal L}'_E = {-i \over 2} d^q_C \bar q \sigma_{\mu\nu} \gamma_5 T^a q
G^{\mu\nu a}
\label{twentyeight}
\ee
and
\be
{\cal L}^{\prime\prime}_E = - {1 \over 6} d_G f_{abc} G_{\mu\nu a}
G^{\nu\rho}_b \tilde G^\mu_{\rho c},
\label{twentynine}
\ee
where $T^a$ are the $SU(3)$ generators, $f_{abc}$ the Gell-Mann
coefficients and $G^{\mu\nu a}$ the gluonic field tensors \cite{b23}.  Their
contributions to the neutron EDM were earlier supposed to be small
with respect to the quark EDM contribution of eq.~(\ref{twentyseven})
\cite{b22}.  But as demonstrated in ref.~\cite{b3}, the large internal cancellation
between the chargino and the gluino contributions to the $d^q$ can make the
net quark EDM contribution comparable to those from the quark
chromoelectric and the gluonic dimension-six operators over certain
regions of parameter space.  In the present analysis we have included
each of these three contributions to the neutron EDM following ref.~\cite{b3}.

We have investigated the ranges of the phase angles $\phi_\mu$ and
$\phi_{A_0}$, over which the predicted electron and neutron EDMs can be
reconciled with the corresponding experimental limits \cite{b9},
\be
d^e < 4.3 \times 10^{-27} \ {\rm ecm}, \ d^n < 6.3 \times 10^{-26} \
{\rm ecm}.
\label{thirty}
\ee
It may be noted here that the above neutron EDM limit is a factor of 2
smaller than that considered in most of the previous analyses
\cite{b3,b5,b6,b7}.  Although it is still an order of magnitude larger than the
electron EDM, both of these provide comparable constraints over the parameter 
space of interest in our analysis.

Figs. (3a), and (3b) show the range of $\phi_\mu$ at $\tan \beta=5$, 
over which the predicted
electron and neutron EDMs can be reconciled with the experimental
limits (\ref{thirty}), by allowing a variation of $\phi_{A_0}$ over the 
range of $-\pi$ to $\pi$.  It should be mentioned here that the
chargino contribution to $d^n$ or $d^e$ overshoots the corresponding
experimental limits for the moderate values of $|\phi_\mu|$ ($\sim 0.1 -
0.2$ radian) shown in Figs. (3a) and (3b).  What helps to satisfy 
the experimental limits is a cancelling contribution from gluino 
exchange for $d^n$
(and neutralino exchange for $d^e$).  Consequently, there is a strong
correlation between the two phase angles, as noted in earlier
analyses.  In particular the maximal allowed value of $|\phi_\mu|$ for a
given $m_0$ and $|A_0|$ corresponds to $|\phi_{A_0}| \sim \pi/2$, and there 
is an opposite sign correlation between the phases.  The
relatively smaller range of $\phi_\mu$ at moderate $|A_0|$ ($< 2000$
GeV) is due to a larger coefficient of $\sin\phi_\mu$ in the chargino
contribution of eq.~(\ref{twentythree}) in comparison with the coefficient of
$\sin\phi_{A_0}$ in the gluino contribution coming from
eqs. (\ref{sixteen},\ref{seventeen},\ref{nineteen}).  It is seen that 
only for very large $|A_0|$ ($\geq 6000$ GeV) the two 
coefficients become comparable; and one 
can satisfy the experimental limits for any value of $\phi_\mu$.  
But, one has to pay a high price in terms of the fine-tuning parameter
amounting to $C > 1000$. Besides, the purely gluonic dimension six operators 
play an effective role in the cancellation mechanism in this region.  

It should be added here that in plotting Figs. (3a) to (3d) 
we have scanned the $m_0$ space in 50 GeV bins.  Decreasing the 
size of this bin further leads to occasional spikes in the
maximum value of $|\phi_\mu|$.  
This is indeed an important effect reflecting further suppression of
the neutron EDM due to internal cancellation amongst the electric
dipole, chromoelectric dipole and the gluonic operator contributions.
As already noted in ref. \cite{b3}, such a drastic suppression of the EDM
can occur over narrow ranges of the SUGRA parameters due to these
cancellations.  But in the present analysis we shall concentrate only
on those results  which hold over wide range of SUGRA parameters,
granting a possible correlation between the two phases.  

In Figs. (3a) and (3b) we have indicated the fine-tuning parameter $C$ 
at some specific points on the fixed $|A_0|$ contours.  Essentially, the
point marked on each contour corresponds to the region of the right 
tip of the fixed $C$ contours 
for $\phi_{A_0}=0$ of Figs. (1a) and (1b).  Comparing 
the two figures one
can easily see that $C$ hardly varies up to
the marked point on each $|A_0|$ contour.  Thus one can accommodate at
least moderate values of $|\phi_\mu^{\rm max}| = 0.1-0.2$ radian for $C =
100 - 200$, i.e. without paying any price in term of fine-tuning.  
Figs.~(3c) and (3d) show the analogous ranges of $\phi_\mu$ at $\tan\beta =
10$.  We see that for given values of $|A_0|$, corresponding to the similar
values of $C$, the $\phi_\mu$ range is less than
half the size of that of Figs.(3a) or (3b).  The underlying 
reason of course is the
comparatively larger coefficients of the chargino contributions for
$d^d$ of eq.~(\ref{twentyone}) and $d^e$ of eq.~(\ref{twentytwo}) at large
$\tan\beta$.  Thus the EDM limits disfavour large values of
$\tan\beta$ ($\geq 10$).

Finally, we also analyse the case of $\tan \beta=3$ and $M_{1/2}=300$~GeV, 
as displayed in Figs.~(4a) and (4b). Fig.~(4a) shows the 
contours for constant fine-tuning $C$ 
which are very different from what we found in Fig. (1a). Such a low value of 
$\tan \beta$ falls outside the focus point scenario \cite{b8}. 
Besides, it is disfavoured by the LEP limit on the lightest Higgs mass \cite{b9}.
Nonetheless most of the previous EDM analyses have
concentrated in this low $\tan\beta$ region \cite{b3,b6,b7}, since it
corresponds to smaller coefficients of the chargino contributions 
of eqs.~(\ref{twentyone}) and (\ref{twentytwo}). However, in this region of 
$\tan \beta$ the fine-tuning parameter $C$ steadily increases with $m_0$ 
unlike what one finds in the focus point scenario.
Here, $C$ is same as $C_{\mu_0}$; 
and a contour of constant $C_{\mu_0}$ is a part of an 
ellipse \cite{b17}. Fig. (4b) shows the variation of maximal 
$|\phi_\mu|$ with $m_0$ for various 
$|A_0|$ values. Unlike Figs. (3a) to (3d), 
$C$ increases here rapidly along the contours of
constant $|A_0|$. 
Consequently, a $|\phi_\mu^{\rm max}|$ of 0.1 radian would correspond 
to a fine-tuning measure $C \simeq 200$, which is larger than 
the value required at $\tan \beta=5$ (Fig. (3a)).
\bigskip

\noindent {\bf Summary:}
\medskip

\nobreak
We have analysed the electron and neutron
EDMs in the focus point scenario of the minimal SUGRA model along with
the fine-tuning parameter.  In this scenario the soft scalar mass 
$m_0$ can go up to 2 TeV without affecting the fine-tuning parameter.
Similarly, the trilinear coupling parameter can be increased from 0 to
1.5 TeV without any appreciable increase in fine-tuning.  The 
large $m_0$ values correspond to large masses for the 1st generation of
sfermions which helps to suppress the one-loop SUSY 
contributions to the EDMs.
Moreover, the large $|A_0|$ corresponds to larger gluino (neutralino)
contribution to quark (electron) EDM, which can cancel the chargino
contribution more effectively.  Therefore, one can satisfy the
experimental limits of the electron and the neutron EDMs without assuming
unnaturally small phases $\phi_\mu$ and $\phi_{A_0}$ 
for $m_0$ and $|A_0|$ values of $\sim 2$ TeV each.  
But this is possible only for a moderate value of $\tan\beta \simeq 5$. 
The chargino contributions to the EDMs increase with
$\tan\beta$, so that the experimental limits cannot be satisfied
without assuming small $|\phi_\mu|$ or a large fine-tuning parameter 
$C$ for $\tan\beta > 10$.  Since the
completion of this work a general phenomenological analysis in the
focus point scenario including the EDMs has appeared recently in 
ref.~\cite{b24}.
However the present work contains a more detailed treatment of this issue.  

U.C. and D.P.R. thank Probir Roy for helpful discussions.

\newpage
\begin{figure}[hbt]
\centerline{\epsfig{file=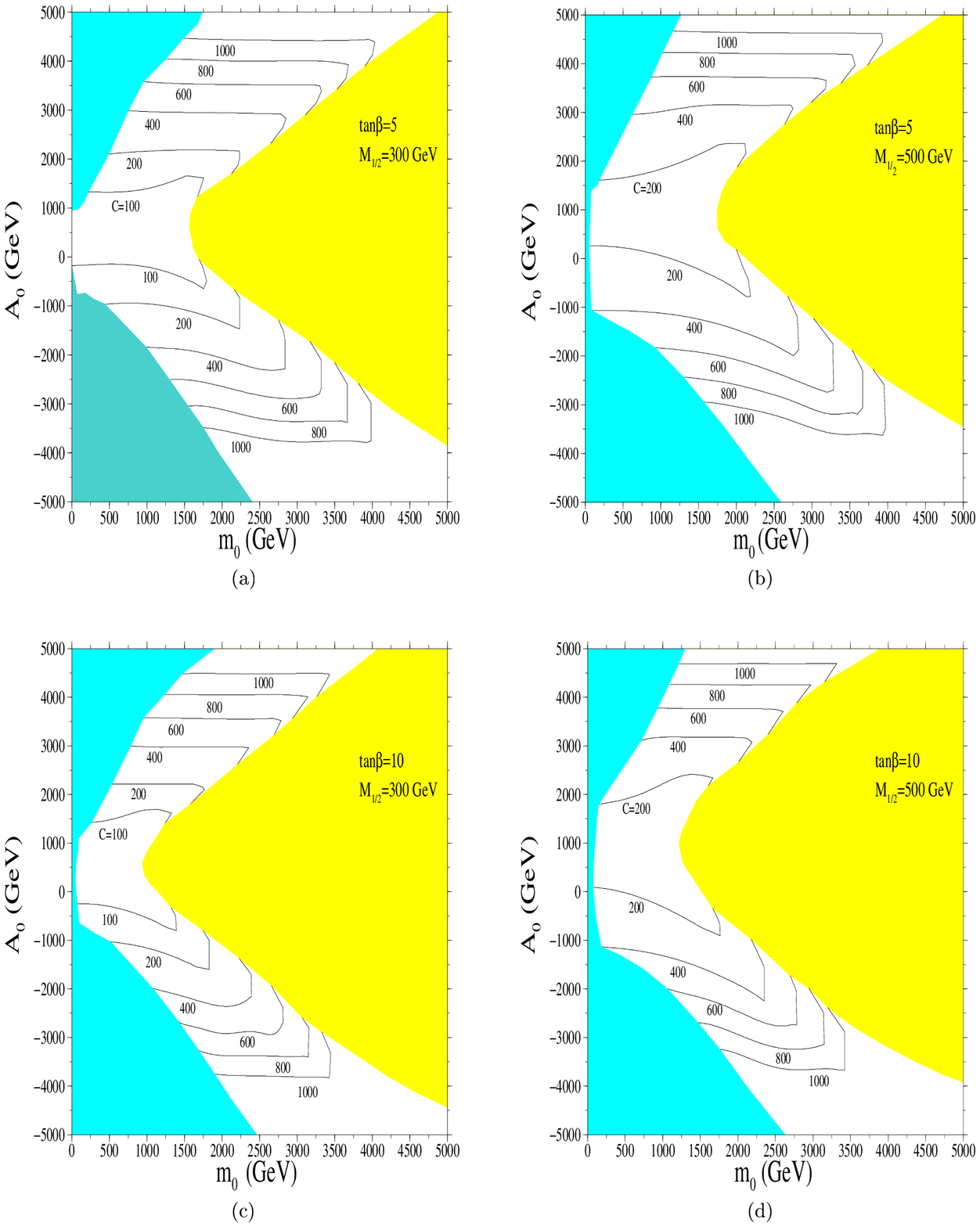,width=15cm}}
\vspace*{-1.0in}
\caption{
Lines of constant fine-tuning $C$ in $(m_0-A_0)$ plane for different values
of $\tan\beta$ and $M_{1/2}$.
Here $\phi_\mu=0$, and for each contours $\phi_{A_0}$ corresponds to 
$0$ and $\pi$ for the upper and the lower parts respectively.  
The shaded areas in the right 
represent the excluded regions due to the chargino mass limit and 
the electroweak radiative breaking constraint. The shaded areas in the left 
are excluded by the top squark mass bounds.}
\label{fine_tune}
\end{figure}
\newpage
\begin{figure}[hbt]
\centerline{\epsfig{file=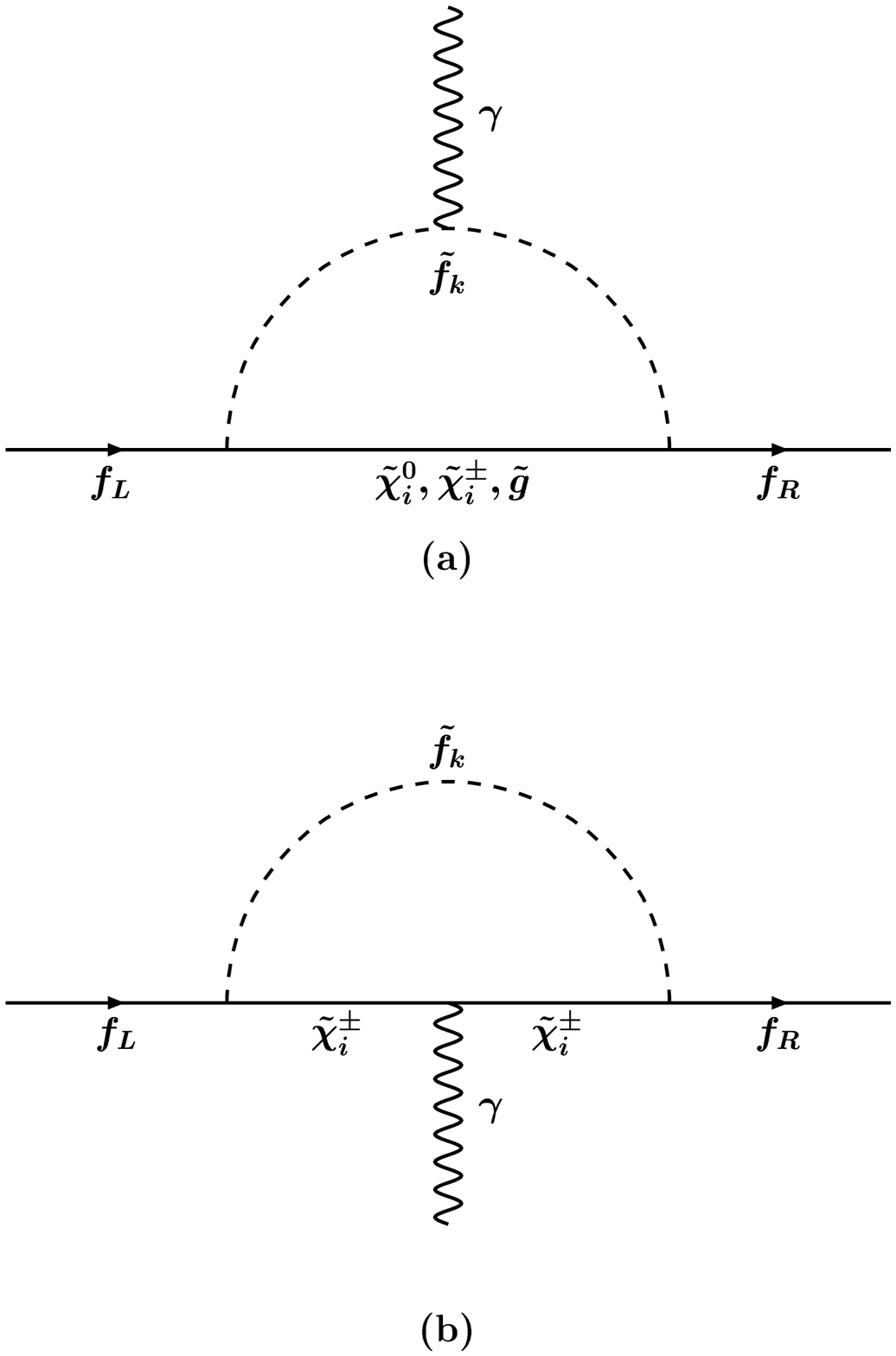,width=15cm}}
\vspace*{-1.0in}
\caption{One loop diagrams contributing to the electric dipole
moments of quarks and leptons.}
\label{Feynman}
\end{figure}

\newpage
\begin{figure}[hbt]
\centerline{\epsfig{file=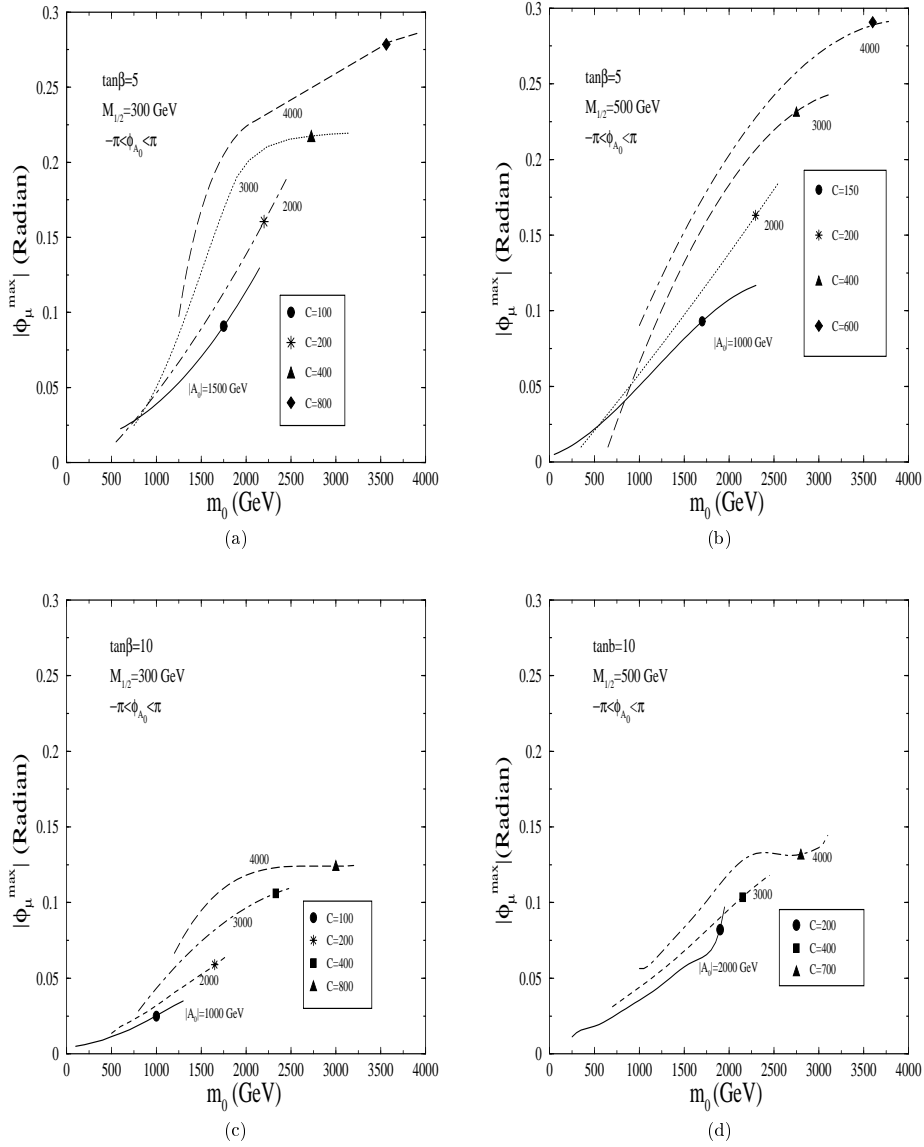,width=15cm}}
\vspace*{-1.0in}
\caption{ 
Lines of constant $| A_0 |$ in the plane of $(|{\phi_\mu}^{max}|-m_0)$ for 
different cases of $\tan \beta$ and $M_{1/2}$. Here the maximum value
of $|\phi_\mu|$ is obtained by varying $\phi_{A_0}$ from $-\pi$ to $\pi$. 
The symbols shown in the figures refer to the appropriate 
values of the fine-tuning measure $C$ which remains practically constant 
on each curve up to the marked point. 
}
\label{phase_graph}
\end{figure}

\newpage
\begin{figure}[hbt]
\centerline{\epsfig{file=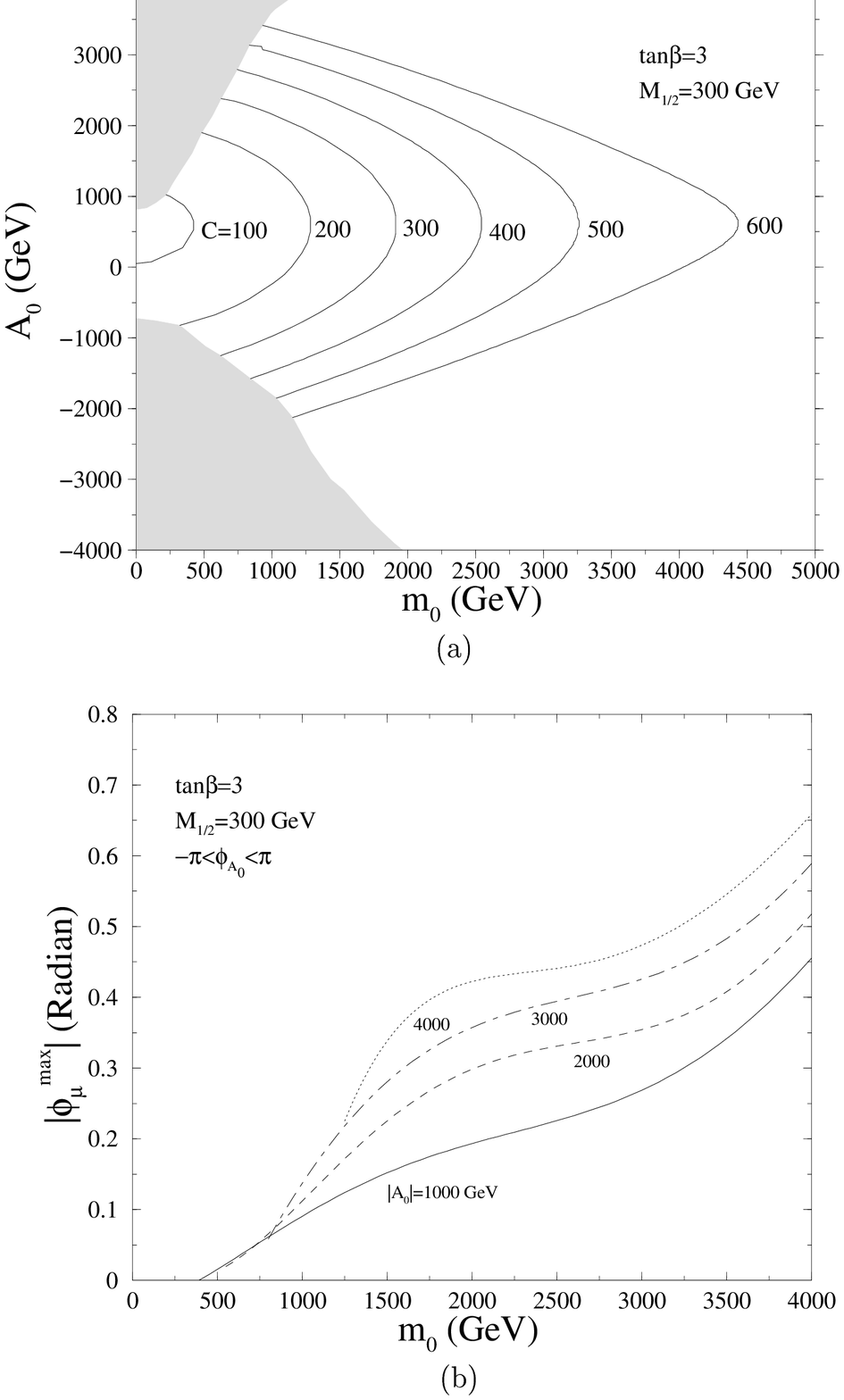,width=15cm}}
\vspace*{-1.0in}
\caption{ 
(a): Lines of constant fine-tuning $C$ in $(m_0-A_0)$ plane for
$\tan\beta=3$ and $M_{1/2}=300$~GeV. 
Here $\phi_\mu=0$, and for each contours $\phi_{A_0}$ corresponds to 
$0$ and $\pi$ for the upper and the lower parts respectively.  
The shaded areas in the left are excluded by the top squark mass bounds.
(b):
Lines of constant $| A_0 |$ in the plane of $|{\phi_\mu}^{max}|-m_0$ for 
$\tan \beta=3$ and $M_{1/2}=300$~GeV. Here the maximum value
of $|\phi_\mu|$ is obtained by varying $\phi_{A_0}$ from $-\pi$ to $\pi$. 
Unlike Figs.(3a) to (3d), here $C$ strongly varies along 
the constant $|A_0|$ contours.}
\label{fine_tune_tan3}
\end{figure}

\end{document}